\documentclass[preprint,showpacs,prc,aps,<floats,superscriptaddress]{revtex4-1}

\begin{document}
\begin{normalsize}

\title{Determination of the number of participating nucleons in p+p interactions at SPS energies}

 \affiliation{Fachbereich Physik der Universit\"{a}t, Frankfurt,
 Germany.}
 
 \author{H.~Str\"{o}bele}
       \affiliation{Fachbereich Physik der Universit\"{a}t, Frankfurt,
 Germany.}

\date{\today}

\begin{abstract}
In the past the analyses of inelastic p+p collision for example in terms of the hadron gas model have assumed that always both protons participate in the interaction. In this article we show that (at $\sqrt{s} = 17.3~GeV$) on average only 1,89 protons are interacting.
Measurements of the mean multiplicities of protons, neutrons, charged pions, charged kaons and $\Lambda$ hyperons allow to compute the number of initial state nucleons using conservation of baryon number, strangeness, and isospin. We further argue that once the number of initial state nucleons is given and the mean multiplicities of protons and charged pions are known in p+p interactions at center-of-mass energies below a few tens of GeV, the yields of net neutrons, kaons, and hyperons can be estimated with a precision of a few percent using again baryon and isospin conservation.
\end{abstract}

 
\maketitle

Mean particle multiplicities are used to characterize the final state of inelastic interactions between elementary hadrons and between nuclei. This applies to total multiplicities as well as to the yields in full phase space of the different hadron species which populate the final state and to energies below $\sqrt{s}$ of a few tens of GeV. Experimentally the determination of the mean particle multiplicities per event is a task with strongly varying complexity. Usually it is possible only for certain particle species. The information provided by experiments can be enhanced using conservation laws. In this article we show that in contrast to common believe the average number of interacting protons (in p+p collisions at a beam energy of 158 GeV/c) is significantly smaller than two. This finding is based on the measurement of the mean multiplicities of protons, neutrons, charged pions, $\Lambda$ hyperons and charged kaons. 

The effective mean number of interacting nucleons $\langle N\rangle$ is equal to the mean net number of baryons in the finals state.
Net means particle minus anti-particle.\\
\begin{equation}
\langle N\rangle = \langle p\rangle - \langle\overline{p}\rangle + \langle n\rangle - \langle\overline{n}\rangle + \langle Y\rangle - \langle\overline{Y}\rangle
\label{Nm}
\end{equation}
We use this equation and the experimental results obtained by the NA49 collaboration on the mean multiplicities of protons and neutrons \cite{f1}, on charged pions \cite{f2}, on charged kaons \cite{f3} and on $\Lambda$  hyperons \cite{Andr} from the NA61 collaboration to calculate  $\langle N\rangle$ for inelastic p+p collisions at a beam momentum of 158 GeV/c.
The values of the first two differences are listed in table \ref{Mult}. The mean number net hyperons is small, and its magnitude can be estimated on the basis of the measured $\Lambda$ + $\Sigma^0$ multiplicity using hyperon multiplicities from a Hadron Gas Model (HGM \cite{Begun}), which are also listed in table \ref{Mult}. We find 

\begin{equation}
\langle Y\rangle - \langle\overline{Y}\rangle) = 1.27\cdot(\langle\Lambda + \Sigma^0\rangle) 
\label{Yref}. 
\end{equation}
With all terms in equation \ref{Nm} known 
we obtain for the mean number of interacting nucleons $\langle N\rangle = 1.87$.\\

Conservation of isospin provides a second method to calculate $\langle N\rangle$.
In the following $\ll i\gg$ stands for the mean value of the third isospin component $I_3$ of particle $i$ (see table \ref{I3}) which has the mean net multiplicity $\langle i\rangle$. 
Isospin conservation provides the following equation in which $\ll N\gg$ denotes the average isospin in the initial state (we mean its third component if we write "isospin") :
\begin{equation}
\ll N\gg = \ll p\gg + \ll n\gg + \ll\pi^+\gg + \ll\pi^-\gg + \ll K\gg + \ll Y\gg 
\label{Ni}
\end{equation}
with
\begin{equation}
\ll K\gg = (1/2)\cdot\langle K^+\rangle - (1/2)\cdot\langle K^-\rangle - (1/2)\cdot\langle K^0\rangle + (1/2)\cdot\langle\overline{K}^0\rangle
\label{K}
\end{equation}
and 
\begin{equation}
\ll Y\gg = \langle\Sigma^+\rangle - \langle\Sigma^-\rangle + (1/2)\cdot\langle\Xi^0\rangle - (1/2)\cdot\langle\Xi^-\rangle
\label{Y}
\end{equation}
The kaon term can be neglected, if the net number of charged kaons equals the net number of neutral kaons, because the sign of $I_3$ of the neutral kaons is the opposite of the one of the charged kaons (see table \ref{I3}). For the time being we also neglect the hyperon term by assuming that the net multiplicities  of $\langle\Sigma^+\rangle$ and $\langle\Sigma^-\rangle$ hyperons are the same. The $\ll\Xi\gg$ contribution is negligibly small at SPS energy (see table \ref{Mult}). Using the experimental values for $\langle p\rangle -  \langle\overline{p}\rangle,  \langle n\rangle - \langle\overline{n}\rangle$ and $\langle \pi^+\rangle, \langle \pi°-\rangle$
listed in table \ref{Mult} yields $\ll N\gg$ = 0.92 and $\langle N\rangle$ = 1.84. 

 \begin{table}
 \begin{ruledtabular}
  \begin{tabular}{ r r r r r r}
		& multiplicities	& & & \\
particle	& measured	 & computed	& HGM	\\
\hline
net proton	& 1.12		&		& 1.00\\
$\pi^+$		& 3.02		&		& 3.16\\
$\pi^-$		& 2.36		&		& 2.37\\
net neutron	& 0.60		 & 0.58		& 0.715\\
net hyperons	&		 & 0.152	& 0.168\\
$K-\overline{K}$&		 & 0.152	& 0.163\\
$K^+$ - $K^-$	& 0.097		 &		& 0.099\\
$K^+$		& 0.227		&		& 0.232\\
$K^-$		& 0.13		&		& 0.133\\
$K^0$		&		&		& 0.206\\
$\overline{K}^0$&		&		& 0.142\\
$K^0_s$		& 0.18		&		& 0.174\\
$\Lambda$	& 0.12		&		& 0.132\\
net $\Lambda$	& 		&		& 0.0935\\
net $\Sigma^0$	& 		&		& 0.0246\\
net $\Sigma^+$	& 		&		& 0.0289\\
net $\Sigma^-$	& 		&		& 0.0171\\
net $\Xi^-$	& 		&		& 0.0015\\
net $\Xi^0$	& 		&		& 0,0018\\
\end{tabular}
 \end{ruledtabular}
\caption{\label{Mult}Experimental, caculated, and HGM \cite{Begun} multiplicities in inelastic p+p interactions at 158 GeV/c. The net neutron multiplicity was calculated by integrating dn/d$x_F$ of neutrons in table 11 of \cite{f1} (respecting the varying binsize) and by subtracting the anti-neutrons using the relation (12) in there. The experimental value for the $\Lambda$ multiplicity contains also the $\Sigma^0$ hyperon multiplicity.} 
\end {table}

The kaon term in equation \ref{Ni} is small but not zero. The unknowns in equation \ref{K} are the multiplicities of the neutral kaons.
These are obtained from strangeness conservation (equation \ref{Lref}) and the measured $\Lambda$ multiplicity, from which the net multiplicity of all hyperons is inferred by equation \ref{Yref}.
\begin{equation}
\langle K^0\rangle - \langle\overline{K}^0\rangle = \langle Y\rangle + \langle\overline{Y}\rangle - \langle K^+\rangle + \langle K^-\rangle 
\label{Lref}
\end{equation}
The multiplicity of net hyperons derived from the $\Lambda$ multiplicity is 0.152. Thus 
$\langle K^0\rangle - \langle\overline{K}^0\rangle = 0.152 - 0.097$ and the isospin in the kaons becomes $(1/2)\cdot(0.097-0.055) = 0.021$.
 
Finally we need to estimate the isospin in the charged $\Sigma$ hyperons, which is the difference of their mean multiplicities. We compute this difference from the corresponding multiplicities provided by the HG model (see table \ref{Mult}) scaled by the ratio of experimental and HGM $\Lambda$ + $\Sigma^0$ multiplicities (0.12/1.32).
The net Isospin in the charged $\Sigma$ hyperons then becomes $(0.0289-0.0171)\cdot(0.12/1.32) = 0.011$. With this the total correction to $\ll N\gg$ equals 0.032. The final result for  $\ll N\gg$ from isospin conservation is 0.952 and for $\langle N\rangle$ 1.904 which is in reasonable agreement with the value obtained from baryon conservation (1.87). Nevertheless this difference suggests to repeat the HGM fit with an an initial proton multiplicity of roughly 1.9 instead of 2.0. In fact the multiplicity of net final state baryons from the HGM calculation (1.883) is also different from the value chosen as initial condition (2.0).

The error on the results obtained for the number of participating nucleons is dominated by the uncertainty of the neutron yield in the final state. It is 6\% of the neutron multiplicity and 0.04 in magnitude. All other experimental uncertainties are of order of 2\%-3\%.
As our final result we take the average value of 1.87 and 1.904, which equals 1.89 $\pm~0.04$.

In summary we have determined the number of participating nucleons in p+p interactions at 158 GeV/c beam momentum. The difference to the commonly employed value of 2.0 is assumed to be due to single diffractive events. Their cross section is roughly 5 mb \cite{sd-dd}, wich is 16\% of the total inelastic cross section. Thus in 16\% of all inelastic events one proton scatters pseudo elastically, which means that the number of really participating nucleons is 1.84 in reasonable agreement with the found value of 1.89. 

The neutron multiplicity is rarely measured in hadronic interactions. If, however, the number of participating nucleons is known in p+p interactions at a specific energy, for example from the diffraction cross section, then the method described in this article can be used to estimate the neutron multiplicity from the one of the protons and the charged pions neglecting the last two terms in equation \ref{Ni}. Then this neutron multiplicity and baryon conservation provide an estimate of the net hyperon ($\langle Y\rangle - \langle\overline{Y}\rangle$) and net kaon ($\langle K^+\rangle - \langle K^-\rangle + \langle K^0\rangle - \langle\overline{K}^0\rangle$) multiplicities. The uncertainty introduced by the omission of the last two terms in equation \ref{Ni} have been calculated above and are of order of 5\% of the neutron multiplicity. Additional multiplicity measurements, for example of charged kaons and $\Lambda$ hyperons will reduce this uncertainty significantly.

In principle the presented scheme can be applied also to nucleus-nucleus  collisions, but the smaller net isospin per nucleon pair and the large number of pions may inroduce too large statistical errors. 

\begin{table}
 \begin{ruledtabular}
  \begin{tabular}{ r r r r r r}
 hyperons 	& $I_3$	 & mesons	& $I_3$ 	& nucleons 	& $I_3$\\
\hline
$\Lambda$	& 0		& $\pi^+$	& +1		& p		& 1/2\\
$\Sigma^+$	& 1		& $\pi^-$	& -1		& n		& -1/2\\
$\Sigma^-$	& -1		& $\pi^0$	&  0\\
$\Sigma^0$	& 0		& $K^+$		& +1/2\\
$\Xi^-$		& -1/2		& $K^-$		& -1/2\\
$\Xi^0$		& +1/2		& $K^0$		& -1/2\\
$\overline{Xi}^+$& +1/2	& $\overline{K}^0$	& +1/2\\
$\overline{Xi}^0$& -1/2\\
$\Omega$	& 0\\
$\overline{\Omega}$&0 \\ 
\end{tabular}
 \end{ruledtabular}

\caption{\label{I3}Third isospin component ($I_3$) of stable hadrons}
\end {table}


\end{normalsize}
\end{document}